\documentclass[prl,reprint,showpacs] {revtex4-1}
\usepackage{amsmath,graphicx}

\usepackage{times}
\usepackage[colorlinks=true,linkcolor=blue,urlcolor=blue,citecolor=blue,breaklinks=true]{hyperref}

\def \dif {{d}}

\date\today
\begin{document}
	
\title{Spectrum of elementary excitations in Galilean-invariant integrable models}
\author{Aleksandra Petkovi\'{c}}
\author{Zoran Ristivojevic}
\affiliation{Laboratoire de Physique Th\'{e}orique, Universit\'{e} de Toulouse, CNRS, UPS, 31062 Toulouse, France}

\begin{abstract}
The spectrum of elementary excitations in one-dimensional quantum liquids is generically linear at low momenta. It is characterized by the sound velocity that can be related to the ground-state energy. Here we study the spectrum at higher momenta in Galilean-invariant integrable models. Somewhat surprisingly, we show that the spectrum at arbitrary momentum is fully determined by the properties of the ground state. We find general exact relations for the coefficients of several terms in the expansion of the excitation energy at low momenta and arbitrary interaction and express them in terms of the Luttinger liquid parameter. We apply the obtained formulas to the Lieb-Liniger model and obtain several new results.
\end{abstract}

\maketitle

The renaissance of studies of interacting bosons has started with  realizations of Bose-Einstein condensates in cold gases \cite{anderson_observation_1995,bradley_evidence_1995,davis_bose-einstein_1995}. Such experiments are carried out in a highly adjustable manner and yielded the observations of numerous quantum phenomena \cite{bloch_many-body_2008}. In particular, the achievement of a one-dimensional system of interacting bosons with controllable correlations  \cite{paredes_tonksgirardeau_2004,kinoshita_observation_2004,kinoshita_quantum_2006,kruger_weakly_2010} have strongly stimulated theoretical activities \cite{cazalilla_one_2011,guan_fermi_2013}. 

One-dimensional interacting systems of quantum particles are qualitatively different from their higher-dimensional counterparts \cite{Giamarchi}. The reason is the strong effect of quantum fluctuations, which renders the ground state in one dimension to be a liquid. At lowest energies, such a quantum state is well described by the phenomenological Luttinger liquid theory \cite{haldane_effective_1981,Giamarchi}. It is characterized by two parameters:One is the excitation velocity, and the other is the so-called Luttinger liquid parameter that controls, e.g., the long-distance decay of the single-particle correlation function. The elementary excitations in a Luttinger liquid have a linear spectrum.

To access the real spectrum of a liquid that is generically nonlinear, one must invoke the approaches that account for the deviations from the Luttinger liquid theory \cite{imambekov_one-dimensional_2012}. At low momenta, the excitation spectrum of elementary excitations of a Galilean-invariant quantum liquid is expected to be of the form \cite{rozhkov_fermionic_2005}
\begin{align}\label{disp}
\varepsilon^\pm_p=v p\pm\frac{p^2}{2m^*}+\frac{\lambda}{6}p^3\pm\frac{\nu}{24}p^4+ \cdots.
\end{align}
Here the two signs refer to to the two types of elementary excitations \cite{lieb_exact_1963b}. The upper (lower) sign refers to the particle (hole) excitation spectrum. By a  well-known thermodynamic argument, the sound velocity $v$ in Eq.~(\ref{disp}) can be related to the derivative of the ground-state energy $E_0$ with respect to the number of particles $N$ as \cite{cazalilla_one_2011,pines+nozieres,lieb_exact_1963b}:
\begin{align}\label{v}
v=\sqrt{\frac{N}{m}\frac{\partial^2 E_0}{\partial N^2}}.
\end{align}  
Here $m$ is the mass of particles. The quadratic term in the spectrum (\ref{disp}) contains the effective mass of elementary excitations $m^*$, which satisfies the relation 
\begin{align}\label{m*}
\frac{m}{m^*}=\partial_n\left(\frac{n}{\sqrt{K}}\right).
\end{align}
Here $K=\pi\hbar n/mv$ denotes the Luttinger liquid parameter, which is related to the sound velocity due to Galilean invariance \cite{haldane_effective_1981,cazalilla_one_2011,imambekov_one-dimensional_2012}. By $n$ we denote the particle density. The expression for the mass of excitations was initially derived phenomenologically in Refs.~\cite{pereira_dynamical_2006,imambekov_one-dimensional_2012}, while it has been recently proved on a microscopic level in Ref.~\cite{matveev_effective_2016}. 

In this work, we use the microscopic theory to study the one-dimensional Galilean invariant quantum liquids. Remarkably, we show that, in the integrable case, the excitation spectrum at any energy is determined only by the properties of the ground state of the liquid. We find that the ground-state energy feature such as the Luttinger liquid parameter (or the sound velocity) fully determines the excitation spectrum. We obtain several analytical results for the spectrum that are valid at an arbitrary interaction. In particular, we find exact relations for the coefficients $\lambda$ and $\nu$ of Eq.~(\ref{disp}). They are expressed in terms of the parameter $K$ and its dependence on the density. We evaluate $\lambda$ and $\nu$ in the special case of the Lieb-Liniger model.

Understanding the excitation spectrum of a quantum Hamiltonian is one of the most fundamental questions in physics. Besides that, in the one-dimensional case, the spectrum is directly related to various exponents of dynamical correlation functions, such as the spectral function and the dynamical structure factor \cite{khodas_dynamics_2007,pereira_exact_2008,imambekov_exact_2008,cheianov_threshold_2008,matveev_spectral_2008,imambekov_phenomenology_2009,pereira_spectral_2009,zvonarev_edge_2009,kamenev_dynamics_2009,imambekov_one-dimensional_2012,shashi_exact_2012}. The latter function has been recently experimentally probed in the Lieb-Liniger model \cite{meinert_probing_2015,fabbri_dynamical_2015}. Our results have therefore a direct application for the analytical calculation of experimentally accessible quantities.

We study the system of interacting quantum particles described by the Hamiltonian
\begin{align}\label{Horiginal}
H=-\frac{\hbar^2}{2m}\sum_{i=1}^N \frac{\partial^2}{\partial x_i^2}+\frac{1}{2}\sum_{i\neq j} V(x_i-x_j).
\end{align}
We consider the thermodynamic limit, when both the number of particles $N$ and the system size $L$ are large, but the particle density $n=N/L$ is kept fixed. In the case of the Lieb-Liniger model \cite{lieb_exact_1963}, Eq.~(\ref{Horiginal}) describes bosons with contact interaction 
$
V(x)=\frac{\hbar^2 c}{m}\delta(x).
$
The parameter $c>0$ describes the repulsion strength.
For the interaction potential
$V(x)=\frac{\hbar^2}{m}\frac{\varLambda(\varLambda-1)\kappa^2}{\sinh^2(\kappa x)},$ Eq.~(\ref{Horiginal}) corresponds to the hyperbolic Calogero-Sutherland model \cite{sutherland}. The parameter  $\kappa$ controls the range of interaction between particles. 

The two models with particular forms of the interparticle interaction are Bethe ansatz integrable \cite{sutherland,korepin1993book,lieb_exact_1963}. This technique enables us to find the exact solution of the many-body problem. Using Bethe ansatz, one finds the set of $N$ rapidities that characterize the many-body wave function. In the ground state, they are compactly packed between the two Fermi points $-Q$ and $Q$. Here, $Q$ is called the Fermi rapidity. It depends on the particle density $n$ and the details of the interaction. Like the Pauli principle that enforces different momenta in a system of fermions, the interaction enforces the spread of rapidities of interacting bosons. In the thermodynamic limit, the density of rapidities $\rho(k,Q)$ in the ground state satisfies the Lieb equation \cite{lieb_exact_1963,sutherland,korepin1993book} 
\begin{align}\label{rho}
\rho(k,Q)+\frac{1}{2\pi}\int_{-Q}^{Q}\dif k' \theta'_+(k-k') \rho(k',Q)=\frac{1}{2\pi},
\end{align} 
where $Q$ is defined by the relation $n=\int_{-Q}^{Q}\dif k\rho(k,Q)$. The kernel $\theta'_+$ is the derivative of the two-particle phase shift. For the Lieb-Liniger model, the phase shift is $\theta_+(k)=-2\arctan(k/c)$, while for the hyperbolic Calogero-Sutherland model $ 
\theta_+(k)=i\ln\left(\frac{\Gamma(1+ik/2\kappa) \Gamma(\varLambda-ik/2\kappa)}{ \Gamma(1-ik/2\kappa) 
\Gamma(\varLambda+ik/2\kappa)}\right)$, where $\Gamma$ denotes the gamma function. For a nonsingular kernel in Eq.~(\ref{rho}), the density of rapidities is a continuous smooth function \footnote{The notable exception where the density of rapidities is discontinuous is the integrable model with the inverse-square potential, which has a singular two-particle phase shift}. We note that the density of rapidities is an even positive function: $\rho(k,Q)=\rho(-k,Q)>0$.

In an excited state, the densely filled ``Fermi sea'' is perturbed. One can distinguish two types of elementary excitations, classified as type I and type II \cite{lieb_exact_1963b}. A type I (particlelike) excitation corresponds to a new set of rapidities, where one promotes one rapidity from $Q$ to a value above $Q$. A type II (holelike) excitation can be seen as the promotion of one rapidity from a value between $-Q$ and $Q$ to the value just above $Q$. The physical momentum of a right-moving $(p>0)$ elementary excitation is given by \cite{lieb_exact_1963b,sutherland,korepin1993book} 
\begin{align}\label{pE}
p(k)=2\pi\hbar \left|\int_Q^k \dif k' \rho(k',Q)\right|,
\end{align}
while the corresponding energy is $\varepsilon(k)=\left|\int_Q^k \dif k' \sigma(k',Q)\right|$. Here, $k>Q$ $(|k|<Q)$ for type I (type II) excitations. The function $\sigma$ that enters the excitation energy obeys another integral equation \cite{sutherland,korepin1993book,lieb_exact_1963b}:
\begin{align}\label{sigma}
\sigma(k,Q)+\frac{1}{2\pi}\int_{-Q}^{Q}\dif k' \theta'_+(k-k') \sigma(k',Q)=\frac{\hbar^2 k}{m}.
\end{align}
Unlike the ground-state energy that is solely determined by the density of rapidities \cite{lieb_exact_1963}, $E_0=(\hbar^2L/2m)\int_{-Q}^{Q}\dif k k^2\rho(k,Q)$, one must calculate the function $\sigma$ defined by Eq.~(\ref{sigma}) to find the excitation spectrum.  

We were able to show that the solution of the Lieb equation (\ref{rho}) is sufficient in order to find the solution of Eq.~(\ref{sigma}). It can be expressed as
\begin{align}\label{sigmak}
\sigma(k,Q)=-\frac{\hbar^2}{2m}\frac{\dif}{\dif k}\int_{|k|}^{Q}\dif k'\frac{\rho(k,k')\tilde n(k')}{\rho^2(k',k')},
\end{align}
where $\tilde n(k)=\int_{-k}^{k}\dif k'\rho(k',k)$. Equation (\ref{sigmak}) is noteworthy and reveals the deep connection between the ground-state properties with the excitations, since it enables us to express the excitation energy as 
\begin{align}\label{eqs}
\varepsilon(k)=\frac{\hbar^2}{2m}\left|\int_Q^{|k|} \dif k'\frac{\rho(k,k')\tilde n(k')}{\rho(k',k')^2}\right|.
\end{align}
Thus, once one knows the density of rapidities in the ground state $\rho(k,Q)$, one can directly infer the whole excitation spectrum, since both $p$ and $\varepsilon$ are expressed in terms of  $\rho(k,Q)$.

Equations for $p$ [Eq.~(\ref{pE})] and $\varepsilon$ [Eq.~(\ref{eqs})] implicitly define the excitation spectrum of the model (\ref{Horiginal}). It acquires the form (\ref{disp}) at low momenta. In the right-hand side of that formula, various powers $p$ have the prefactors that can be expressed as derivatives of $\varepsilon$ with respect to $p$. Since low $p$ corresponds to $k\to Q$, it becomes evident that the prefactors in Eq.~(\ref{disp}) can be expressed in terms of $\rho(Q,Q)$ and its partial derivatives. For simplicity, let us first consider type I excitations, $k>Q$, and, therefore, the absolute values in the momentum $p(k)$ [Eq.~(\ref{pE})] and the energy $\varepsilon(k)$ [Eq.~(\ref{eqs})] can be omitted. The momentum and the energy are analytic functions and can be expanded into a Taylor series at $k\to Q$. Eliminating $k$ from them yields the spectrum of the form $\varepsilon^+_p$ of Eq.~(\ref{disp}). In the following, we evaluate $v$, $1/m^*$, $\lambda$, and $\nu$.

A formal evaluation of the sound velocity $v=\frac{\partial\varepsilon(k)}{\partial p(k)}\big|_{k=Q}$ yields the expression $v=\hbar n/4\pi m\rho^2(Q,Q)$, since $\tilde n(Q)=n$. On the other hand, for a Galilean-invariant system, the velocity can be also expressed as $v=\pi\hbar n/mK$, leading to $\rho(Q,Q)=\sqrt{K}/2\pi$ \cite{sutherland,korepin1993book}.

The mass of elementary excitations is defined by the relation $1/m^*=\frac{\partial^2\varepsilon(k)}{\partial p(k)^2}\big|_{k=Q}$. From Eqs.~(\ref{pE}) and (\ref{eqs}), one directly obtains  $m/m^*=[1/8\pi^2\rho^2(Q,Q)]\dif\left[{n}/{\rho(Q,Q)}\right] /\dif Q$. Using the derivative of the density with respect to the Fermi rapidity given by \cite{korepin1993book,matveev_effective_2016} $\dif n/\dif Q=K/\pi$, we transform the derivative to be with respect to $n$ and eventually obtain Eq.~(\ref{m*}). We have therefore confirmed the result for the effective mass of Refs.~\cite{pereira_dynamical_2006,imambekov_one-dimensional_2012,matveev_effective_2016} using the microscopic approach.

We now evaluate the cubic term $\lambda=\frac{\partial^3\varepsilon(k)}{\partial p(k)^3}\big|_{k=Q}$ in the spectrum (\ref{disp}). To achieve that, we use the relation
\begin{align}\label{rho''}
\rho''_{QQ}(k,Q)-\rho''_{kk}(k,Q)=2\frac{\dif}{\dif Q}[\ln \rho(Q,Q)]\rho'_Q(k,Q),
\end{align}
which is straightforwardly obtained from the Lieb equation (\ref{rho}). Equation (\ref{rho''}) helps us to reduce the order of derivatives. We also employ the expression for the total derivative of $\rho(Q,Q)$. After an elementary calculation, we obtain
\begin{align}\label{lambda1}
\lambda=\frac{2\pi \left[2\rho'_Q(Q,Q)-\frac{\dif }{\dif Q}\rho(Q,Q)\right]}{\hbar K m^*}+\frac{\sqrt{K} \partial_n \left(\frac{1}{m^*}\right)}{2\pi\hbar}.
\end{align}
Here $\rho'_Q(Q,Q)=\partial \rho(k,Q)/\partial Q|_{k=Q}$. Suitably multiplying Eq.~(\ref{lambda1}) by $m^*\sqrt{K}$ and taking the derivative with respect to $Q$, the partial derivatives in the right hand side are canceled as a consequence of Eq.~(\ref{rho''}). Finally, expressing all the derivatives to be with respect to the density, we find
\begin{align}\label{lambda}
\partial_n (\lambda m^*\sqrt{K})=\frac{1}{\pi\hbar} (\partial_n \sqrt{K})^2-\frac{1}{2\pi\hbar}\partial_n\!\left(\!K \frac{\partial_n m^*}{m^*}\right).
\end{align}
The expression (\ref{lambda}) describes the cubic term in the spectrum (\ref{disp}). It is an exact result valid at any interaction strength. We note that the details of the interaction potential are not important for its derivation once the two-particle scattering phase shift is a nonsingular function. We point out that Eq.~(\ref{lambda}) can be expressed only in terms of the dependence of the Luttinger liquid parameter $K$ on the density and the density itself [cf. Eq.~(\ref{m*})]. Generally, $K$ can be obtained from the ground-state energy using Eq.~(\ref{v}) and its connection to the sound velocity.

The evaluation of $\nu=\frac{\partial^4\varepsilon(k)}{\partial p(k)^4}\big|_{k=Q}$ is more tedious. However, one can proceed along the same lines as for $\lambda$ and use Eq.~(\ref{rho''}) and its derivatives to 
transform various partial derivatives. The final result takes the form
\begin{align}\label{nu}
&\frac{2\pi\hbar}{m^*K^{3/2}}\partial_n(\nu m^* K)=\partial_n^2\lambda-\left(\frac{3 \partial_n v}{v}-\frac{5\partial_n {m^*}}{2{m^*}}\right)\partial_n\lambda\notag\\
&\qquad+\frac{3}{2} \left[\frac{5(\partial_n K)^2}{2K^2}-\frac{\partial_n(n^2\partial_n K)}{n^2 K}+\frac{\partial_n\left(n\partial_n\frac{1}{m^*}\right)}{n\frac{1}{m^*}}\right]\lambda\notag\\
&\qquad-\frac{3}{2v {m^*}^{5/2}}\partial_n\left(\frac{\partial_n m^*}{\sqrt{m^*}}\right).
\end{align}
Equations (\ref{lambda}) and (\ref{nu}) apply to integrable models with nonsingular phase shifts, including the Lieb-Liniger and the hyperbolic Calogero-Sutherland models. They have the form of first-order differential equations and can be solved up to a numerical factor, as we demonstrate below. 

Let us evaluate $\lambda$ and $\nu$ for the Lieb-Liniger model. We characterize the interaction strength by the dimensionless parameter \cite{lieb_exact_1963} $\gamma=c/n$. Solving Eq.~(\ref{lambda}), with the help of the dimensional analysis, we obtain
\begin{align}\label{lambdasol}
\lambda=\frac{\int \dif n (\partial_n \sqrt{K})^2}{\pi\hbar m^*\sqrt{K}} + \frac{\sqrt{K} \partial_n \left(\frac{1}{m^*}\right)}{2\pi\hbar}+\frac{A}{\hbar m^*\sqrt{K}\gamma n},
\end{align}
where $A$ is a numerical constant. The last term of Eq.~(\ref{lambdasol}) accounts for the integration constant. Similarly, for $\nu$ we find
\begin{align}
\nu=\frac{\int\dif n m^* K^{3/2}\,g(n)}{2\pi\hbar m^*K}+\frac{B}{\hbar^2 m^* K\gamma^2 n^2},
\end{align}
where $B$ is a numerical constant, while $g(n)$ denotes the right-hand side of Eq.~(\ref{nu}). The Luttinger liquid parameter $K$ that enters the previous equations can be found from the ground-state energy $E_0$. The latter can be expressed \cite{lieb_exact_1963,popov_theory_1977}  in terms of the dimensionless function $e(\gamma)$ through the relation $E_0=(\hbar^2n^2N/2m)e(\gamma)$. Using Eq.~(\ref{v}) and $K=\pi\hbar n/mv$, one can find $K={\pi}/{\sqrt{3e-2\gamma \frac{\dif e}{\dif\gamma} +\frac{1}{2}\gamma^2\frac{\dif^2 e}{\dif\gamma^2}}}$. 

In the limit of weak interaction, $\gamma\ll 1$, the function $e(\gamma)$ is analytically calculated in Refs.~\cite{popov_theory_1977,tracy_ground_2016,prolhac_ground_2017}. It leads to
\begin{align}\label{K}
K=\frac{\pi}{\sqrt{\gamma}}\left[1+\frac{\sqrt{\gamma}}{4\pi}+\frac{3{\gamma}} {32\pi^2}+\mathcal{O}(\gamma^{3/2})\right].
\end{align}
In order to find the numerical constant $A$ of Eq.~(\ref{lambdasol}) it is sufficient to calculate the leading order result $\rho'_Q(Q,Q)=\sqrt{\pi}/12n\gamma^{5/4}$ in Eq.~(\ref{lambda1}). We have obtained the latter expression using the Wiener-Hopf technique to solve the approximate version of the integral equation (\ref{rho}) that correctly describes $\rho(k,Q)$ in the leading order in $\gamma$ and at $k$ in the vicinity of $Q$  \cite{hutson_circular_1963,popov_theory_1977}. The final result takes the form
\begin{align}\label{lambdaweak}
\lambda=\frac{1}{4\hbar mn\sqrt{\gamma}}\left[1- \frac{11\sqrt{\gamma}}{12\pi}+\frac{19\gamma}{192\pi^2} +\mathcal{O}(\gamma^{3/2})\right].
\end{align}
At the leading order, Eq.~(\ref{lambdaweak}) is in agreement with the result of Ref.~\cite{pustilnik_low-energy_2014}. However, our result (\ref{lambdasol}) is exact and enables us to find all higher-order corrections in the expression for $\lambda$; the first two are given in Eq.~(\ref{lambdaweak}). The evaluation of $\nu$ is done in a similar manner, with the result
\begin{align}\label{nuweak}
\!\!\nu=\frac{5\sqrt{\pi}}{32\hbar^2 n^2m\gamma^{5/4}}\! \left[1-\frac{49\sqrt{\gamma}}{24\pi}+\frac{203\gamma}{128\pi^2} +\mathcal{O}(\gamma^{3/2})\right].\!\!
\end{align}
The results (\ref{lambdaweak}) and (\ref{nuweak}) are valid at weak interaction, $\gamma\ll 1$. They diverge at $\gamma\to 0$. However, the spectrum (\ref{disp}) is applicable for low momenta, $|p|\ll \hbar n\gamma^{3/4}$, and remains finite at $\gamma\to 0$. We note that at higher momenta the spectrum of the type I elementary excitations takes the Bogoliubov form, while the type II excitations correspond to the dark soliton solution of the time-dependent Gross-Pitaevskii equation \cite{kulish_comparison_1976}. In the noninteracting limit one then obtains the usual form of the spectrum $p^2/2m$.

\begin{figure}
	\includegraphics[width=0.9\columnwidth]{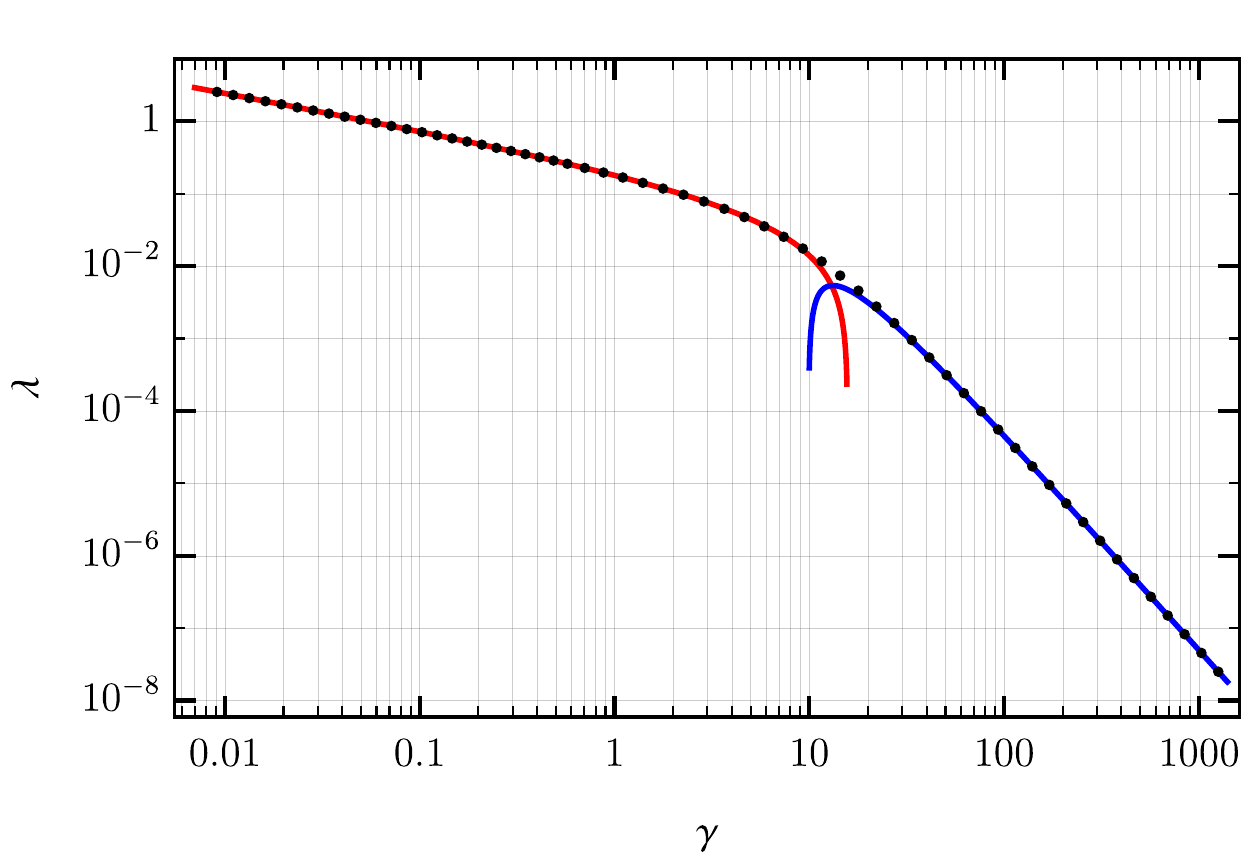}
	\caption{Plot of the coefficient $\lambda$ (in units where $\hbar= n=m=1$) as a function of $\gamma$. The dots represent the numerically exact result, while the two lines are our analytical results given by Eqs.~(\ref{lambdaweak}) and (\ref{lambdastrong}).}\label{fig1}
\end{figure}

\begin{figure}
	\includegraphics[width=0.9\columnwidth]{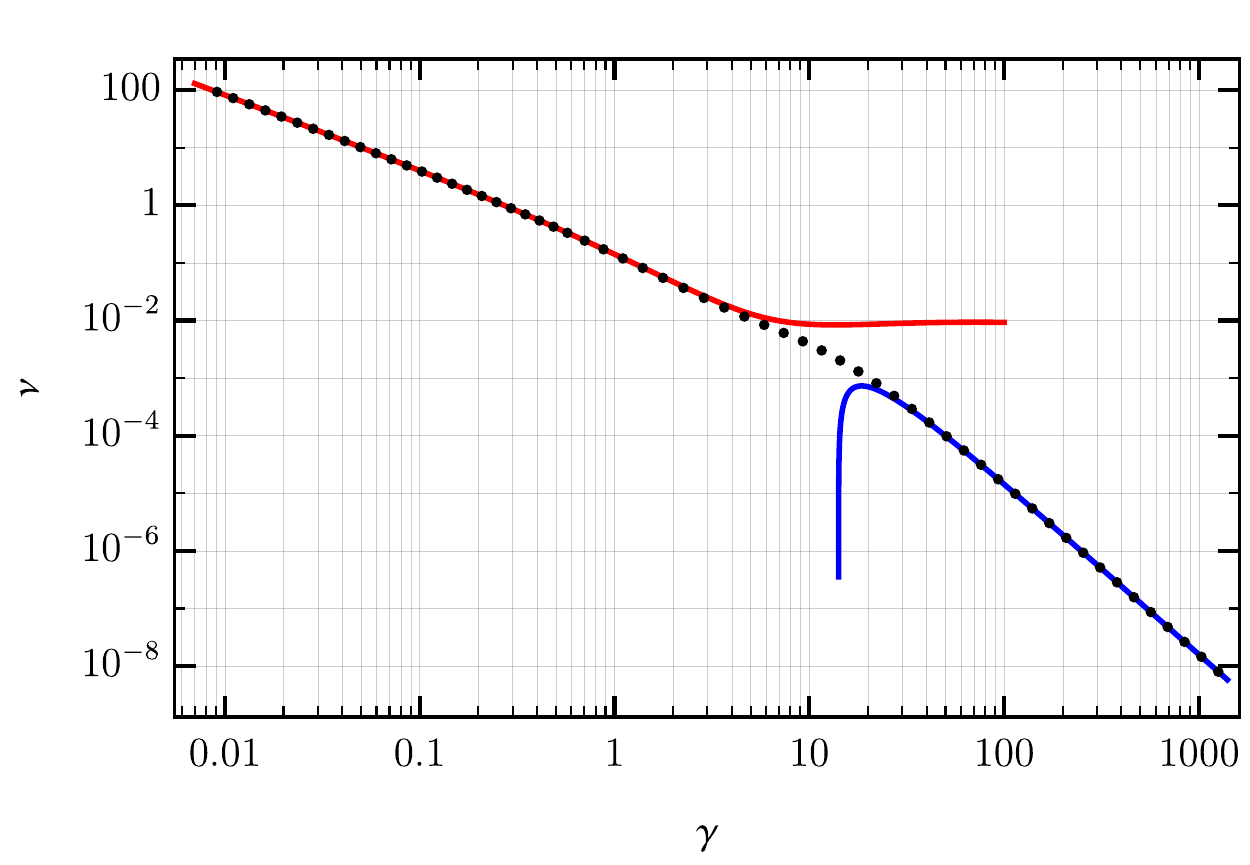}
	\caption{Plot of the coefficient $\nu$ (in units where $\hbar= n=m=1$) as a function of $\gamma$. The dots represent the numerically exact result, while the two lines are our analytical results given by Eqs.~(\ref{nuweak}) and (\ref{nustrong}).}\label{fig2}
\end{figure}

At strong interaction, $\gamma\gg 1$, the kernel of the integral equation (\ref{rho}) can be expanded into a power series and then Eq.~(\ref{rho}) can be solved \cite{lieb_exact_1963,ristivojevic_excitation_2014,lang_ground-state_2017}. For the derivative of the density of rapidities that enter Eq.~(\ref{lambda1}) we find
\begin{align}\label{rhopr}
\rho'_Q(Q,Q)=
\frac{1}{\pi^2 n\gamma}\biggl[1+\frac{4}{\gamma}-\frac{2(\pi^2-2)}{\gamma^2} +\mathcal{O}(\gamma^{-3})
\biggr].
\end{align}
Together with the Luttinger liquid parameter calculated in Ref.~\cite{ristivojevic_excitation_2014}, $K=1+4/\gamma+4/\gamma^2-16\pi^2/3\gamma^3+32\pi^2/3\gamma^4+64\pi^2(-5+3\pi^2)/15\gamma^5+\mathcal{O}(\gamma^{-6})$, Eq.~(\ref{rhopr}) enables us to find the constant in Eq.~(\ref{lambdasol}). We then obtain
\begin{align}\label{lambdastrong}
\lambda=\frac{16\pi}{\hbar n m \gamma^3}\left[1-\frac{10}{\gamma}+\frac{6(10-\pi^2)}{\gamma^2}+\mathcal{O}(\gamma^{-3})\right].
\end{align}
Similarly, we find the expression 
\begin{align}\label{nustrong}
\nu=\frac{16}{\hbar^2 n^2 m\gamma^3}\left[1-\frac{10}{\gamma}-\frac{12(\pi^2-5)}{\gamma^2}+\mathcal{O}(\gamma^{-3})\right].
\end{align}
At the two lowest orders, Eqs.~(\ref{lambdastrong}) and (\ref{nustrong}) are in agreement with Ref.~\cite{ristivojevic_excitation_2014}. In Figs.~\ref{fig1} and \ref{fig2}, we plot the numerical results for $\lambda$ and $\nu$ obtained by solving the Bethe ansatz equations and find perfect agreement with our analytical asymptotic results.

As a by-product of the previous calculation, we can obtain the exact information about the partial derivatives of $\rho(Q,Q)$. In particular, from Eqs.~(\ref{lambda1}) and (\ref{lambda}) we obtain the differential equation showing that the derivative of the density of rapidities at the Fermi rapidity, $\rho'_Q(Q,Q)=\partial \rho(k,Q)/\partial Q|_{k=Q}$, must satisfy:
\begin{align}\label{eqrho2}
\partial_n\left[\frac{8\pi^2}{\sqrt K}\rho'_Q(Q,Q)\right]=\partial_n^2 K+2(\partial_n\sqrt{K})^2.
\end{align}
Equation (\ref{eqrho2}) is exact and valid for a general Galilean invariant integrable model. We note that Eq.~(\ref{eqrho2}) can also be obtained directly from Eq.~(\ref{rho''}). In the special case of the Lieb-Liniger model, the solution of Eq.~(\ref{eqrho2}) follows from Eqs.~(\ref{lambda1}), (\ref{lambdaweak}), and (\ref{lambdastrong}). At weak interaction, $\gamma\ll 1$, we obtain
\begin{align}
\!\!\!\rho'_Q(Q,Q)=
\frac{\sqrt{\pi}}{12n\gamma^{5/4}}\biggl[1+\frac{\sqrt\gamma}{2\pi}+\frac{17\gamma}{128\pi^2}+\mathcal{O}(\gamma^{3/2})\biggr],
\end{align}
while at strong interaction, $\gamma\gg 1$, it is given by Eq.~(\ref{rhopr}). The other partial derivative can now be easily found: $\rho'_k(Q,Q)=\dif\rho(Q,Q)/\dif Q-\rho'_Q(Q,Q)$. In a similar fashion one can obtain second partial derivatives of $\rho(Q,Q)$ from $\nu$; however we do not pursue them here.

Equations (\ref{lambda}) and (\ref{nu}) contain the relations for the cubic and quartic terms in the spectrum (\ref{disp}) at an arbitrary interaction. For the case of the Lieb-Liniger model, the corresponding relations are given by Eqs.~(\ref{lambdaweak}) and  
(\ref{nuweak}) at a weak interaction, while at a strong interaction they are given by Eqs.~(\ref{lambdastrong}) and (\ref{nustrong}). They apply for both cases of type I and type II excitations. Namely, the spectrum of type II excitations, $\varepsilon_p^-$, is defined by Eqs.~(\ref{pE}) and (\ref{eqs}) at $|k|<Q$. In the case of analytic $p(k)$ and $\varepsilon(k)$, it can be formally related by the spectrum $\varepsilon_p^+$ of type I excitations as $\varepsilon_p^-=-\varepsilon_{-p}^+$ \cite{ristivojevic_excitation_2014}. Hence, it is justified to assume the energy spectrum in the form given by Eq.~(\ref{disp}). Moreover, all the relations that we found for the prefactors in Eq.~(\ref{disp}) are equally valid for both types of elementary excitations.

Our results have a direct application for the calculation of exponents in the dynamical correlation functions. The exponents can be related to the derivatives of the excitation spectrum (\ref{disp}), as shown and detailed in Refs.~\cite{imambekov_phenomenology_2009,imambekov_one-dimensional_2012}. In general, they are momentum dependent; such behavior is controlled by the cubic and higher-order terms in the spectrum, which are found in this work. Moreover, our results have an application for the evaluation of the so-called shift function of the integrable model \cite{korepin1993book,imambekov_one-dimensional_2012}, which satisfies an equation of the form (\ref{rho}) where instead of the constant $1/2\pi$, on the right-hand side one has the two-particle phase shift. As was initially shown in Refs.~\cite{pereira_exact_2008,imambekov_exact_2008,cheianov_threshold_2008}, the exponents in the dynamical correlation functions can be (also) related to the shift function, and thus one can use the explicit results for the spectrum to calculate the shift function.

In conclusion, we showed that the spectrum of elementary excitations in one-dimensional Galilean invariant integrable models (with a nonsingular two-particle phase shift) is controlled by the ground-state properties of the system. We found general exact relations for the cubic and quartic terms in the spectrum (\ref{disp}). For the Lieb-Liniger model we extracted the explicit results in the regimes of weak and strong interaction. Our results show that, like the excitation energy, the exponents in the dynamical correlation functions are controlled by the features of the ground state of the liquid. The approach developed here can be used to find higher-order terms in Eq.~(\ref{disp}). It has potential generalizations to the integrable one-dimensional models without Galilean invariance and the multicomponent models. There one would also expect the excitation spectrum to be controlled by the ground-state properties of the system.

We thank K. Matveev for useful discussions.


\begin{thebibliography}{42}%
	\makeatletter
	\providecommand \@ifxundefined [1]{%
		\@ifx{#1\undefined}
	}%
	\providecommand \@ifnum [1]{%
		\ifnum #1\expandafter \@firstoftwo
		\else \expandafter \@secondoftwo
		\fi
	}%
	\providecommand \@ifx [1]{%
		\ifx #1\expandafter \@firstoftwo
		\else \expandafter \@secondoftwo
		\fi
	}%
	\providecommand \natexlab [1]{#1}%
	\providecommand \enquote  [1]{``#1''}%
	\providecommand \bibnamefont  [1]{#1}%
	\providecommand \bibfnamefont [1]{#1}%
	\providecommand \citenamefont [1]{#1}%
	\providecommand \href@noop [0]{\@secondoftwo}%
	\providecommand \href [0]{\begingroup \@sanitize@url \@href}%
	\providecommand \@href[1]{\@@startlink{#1}\@@href}%
	\providecommand \@@href[1]{\endgroup#1\@@endlink}%
	\providecommand \@sanitize@url [0]{\catcode `\\12\catcode `\$12\catcode
		`\&12\catcode `\#12\catcode `\^12\catcode `\_12\catcode `\%12\relax}%
	\providecommand \@@startlink[1]{}%
	\providecommand \@@endlink[0]{}%
	\providecommand \url  [0]{\begingroup\@sanitize@url \@url }%
	\providecommand \@url [1]{\endgroup\@href {#1}{\urlprefix }}%
	\providecommand \urlprefix  [0]{URL }%
	\providecommand \Eprint [0]{\href }%
	\providecommand \doibase [0]{http://dx.doi.org/}%
	\providecommand \selectlanguage [0]{\@gobble}%
	\providecommand \bibinfo  [0]{\@secondoftwo}%
	\providecommand \bibfield  [0]{\@secondoftwo}%
	\providecommand \translation [1]{[#1]}%
	\providecommand \BibitemOpen [0]{}%
	\providecommand \bibitemStop [0]{}%
	\providecommand \bibitemNoStop [0]{.\EOS\space}%
	\providecommand \EOS [0]{\spacefactor3000\relax}%
	\providecommand \BibitemShut  [1]{\csname bibitem#1\endcsname}%
	\let\auto@bib@innerbib\@empty
	\bibitem [{\citenamefont {Anderson}\ \emph {et~al.}(1995)\citenamefont
		{Anderson}, \citenamefont {Ensher}, \citenamefont {Matthews}, \citenamefont
		{Wieman},\ and\ \citenamefont {Cornell}}]{anderson_observation_1995}%
	\BibitemOpen
	\bibfield  {author} {\bibinfo {author} {\bibfnamefont {M.~H.}\ \bibnamefont
			{Anderson}}, \bibinfo {author} {\bibfnamefont {J.~R.}\ \bibnamefont
			{Ensher}}, \bibinfo {author} {\bibfnamefont {M.~R.}\ \bibnamefont
			{Matthews}}, \bibinfo {author} {\bibfnamefont {C.~E.}\ \bibnamefont
			{Wieman}}, \ and\ \bibinfo {author} {\bibfnamefont {E.~A.}\ \bibnamefont
			{Cornell}},\ }\href {\doibase 10.1126/science.269.5221.198} {\bibfield
		{journal} {\bibinfo  {journal} {Science}\ }\textbf {\bibinfo {volume}
			{269}},\ \bibinfo {pages} {198} (\bibinfo {year} {1995})}\BibitemShut
	{NoStop}%
	\bibitem [{\citenamefont {Bradley}\ \emph {et~al.}(1995)\citenamefont
		{Bradley}, \citenamefont {Sackett}, \citenamefont {Tollett},\ and\
		\citenamefont {Hulet}}]{bradley_evidence_1995}%
	\BibitemOpen
	\bibfield  {author} {\bibinfo {author} {\bibfnamefont {C.~C.}\ \bibnamefont
			{Bradley}}, \bibinfo {author} {\bibfnamefont {C.~A.}\ \bibnamefont
			{Sackett}}, \bibinfo {author} {\bibfnamefont {J.~J.}\ \bibnamefont
			{Tollett}}, \ and\ \bibinfo {author} {\bibfnamefont {R.~G.}\ \bibnamefont
			{Hulet}},\ }\href {\doibase 10.1103/PhysRevLett.75.1687} {\bibfield
		{journal} {\bibinfo  {journal} {Phys. Rev. Lett.}\ }\textbf {\bibinfo
			{volume} {75}},\ \bibinfo {pages} {1687} (\bibinfo {year}
		{1995})}\BibitemShut {NoStop}%
	\bibitem [{\citenamefont {Davis}\ \emph {et~al.}(1995)\citenamefont {Davis},
		\citenamefont {Mewes}, \citenamefont {Andrews}, \citenamefont {van Druten},
		\citenamefont {Durfee}, \citenamefont {Kurn},\ and\ \citenamefont
		{Ketterle}}]{davis_bose-einstein_1995}%
	\BibitemOpen
	\bibfield  {author} {\bibinfo {author} {\bibfnamefont {K.~B.}\ \bibnamefont
			{Davis}}, \bibinfo {author} {\bibfnamefont {M.~O.}\ \bibnamefont {Mewes}},
		\bibinfo {author} {\bibfnamefont {M.~R.}\ \bibnamefont {Andrews}}, \bibinfo
		{author} {\bibfnamefont {N.~J.}\ \bibnamefont {van Druten}}, \bibinfo
		{author} {\bibfnamefont {D.~S.}\ \bibnamefont {Durfee}}, \bibinfo {author}
		{\bibfnamefont {D.~M.}\ \bibnamefont {Kurn}}, \ and\ \bibinfo {author}
		{\bibfnamefont {W.}~\bibnamefont {Ketterle}},\ }\href {\doibase
		10.1103/PhysRevLett.75.3969} {\bibfield  {journal} {\bibinfo  {journal}
			{Phys. Rev. Lett.}\ }\textbf {\bibinfo {volume} {75}},\ \bibinfo {pages}
		{3969} (\bibinfo {year} {1995})}\BibitemShut {NoStop}%
	\bibitem [{\citenamefont {Bloch}\ \emph {et~al.}(2008)\citenamefont {Bloch},
		\citenamefont {Dalibard},\ and\ \citenamefont
		{Zwerger}}]{bloch_many-body_2008}%
	\BibitemOpen
	\bibfield  {author} {\bibinfo {author} {\bibfnamefont {I.}~\bibnamefont
			{Bloch}}, \bibinfo {author} {\bibfnamefont {J.}~\bibnamefont {Dalibard}}, \
		and\ \bibinfo {author} {\bibfnamefont {W.}~\bibnamefont {Zwerger}},\ }\href
	{\doibase 10.1103/RevModPhys.80.885} {\bibfield  {journal} {\bibinfo
			{journal} {Rev. Mod. Phys.}\ }\textbf {\bibinfo {volume} {80}},\ \bibinfo
		{pages} {885} (\bibinfo {year} {2008})}\BibitemShut {NoStop}%
	\bibitem [{\citenamefont {Paredes}\ \emph {et~al.}(2004)\citenamefont
		{Paredes}, \citenamefont {Widera}, \citenamefont {Murg}, \citenamefont
		{Mandel}, \citenamefont {F\"olling}, \citenamefont {Cirac}, \citenamefont
		{Shlyapnikov}, \citenamefont {H\"ansch},\ and\ \citenamefont
		{Bloch}}]{paredes_tonksgirardeau_2004}%
	\BibitemOpen
	\bibfield  {author} {\bibinfo {author} {\bibfnamefont {B.}~\bibnamefont
			{Paredes}}, \bibinfo {author} {\bibfnamefont {A.}~\bibnamefont {Widera}},
		\bibinfo {author} {\bibfnamefont {V.}~\bibnamefont {Murg}}, \bibinfo {author}
		{\bibfnamefont {O.}~\bibnamefont {Mandel}}, \bibinfo {author} {\bibfnamefont
			{S.}~\bibnamefont {F\"olling}}, \bibinfo {author} {\bibfnamefont
			{I.}~\bibnamefont {Cirac}}, \bibinfo {author} {\bibfnamefont {G.~V.}\
			\bibnamefont {Shlyapnikov}}, \bibinfo {author} {\bibfnamefont {T.~W.}\
			\bibnamefont {H\"ansch}}, \ and\ \bibinfo {author} {\bibfnamefont
			{I.}~\bibnamefont {Bloch}},\ }\href {\doibase 10.1038/nature02530} {\bibfield
		{journal} {\bibinfo  {journal} {Nature}\ }\textbf {\bibinfo {volume}
			{429}},\ \bibinfo {pages} {277} (\bibinfo {year} {2004})}\BibitemShut
	{NoStop}%
	\bibitem [{\citenamefont {Kinoshita}\ \emph {et~al.}(2004)\citenamefont
		{Kinoshita}, \citenamefont {Wenger},\ and\ \citenamefont
		{Weiss}}]{kinoshita_observation_2004}%
	\BibitemOpen
	\bibfield  {author} {\bibinfo {author} {\bibfnamefont {T.}~\bibnamefont
			{Kinoshita}}, \bibinfo {author} {\bibfnamefont {T.}~\bibnamefont {Wenger}}, \
		and\ \bibinfo {author} {\bibfnamefont {D.~S.}\ \bibnamefont {Weiss}},\ }\href
	{\doibase 10.1126/science.1100700} {\bibfield  {journal} {\bibinfo  {journal}
			{Science}\ }\textbf {\bibinfo {volume} {305}},\ \bibinfo {pages} {1125}
		(\bibinfo {year} {2004})}\BibitemShut {NoStop}%
	\bibitem [{\citenamefont {Kinoshita}\ \emph {et~al.}(2006)\citenamefont
		{Kinoshita}, \citenamefont {Wenger},\ and\ \citenamefont
		{Weiss}}]{kinoshita_quantum_2006}%
	\BibitemOpen
	\bibfield  {author} {\bibinfo {author} {\bibfnamefont {T.}~\bibnamefont
			{Kinoshita}}, \bibinfo {author} {\bibfnamefont {T.}~\bibnamefont {Wenger}}, \
		and\ \bibinfo {author} {\bibfnamefont {D.~S.}\ \bibnamefont {Weiss}},\ }\href
	{\doibase 10.1038/nature04693} {\bibfield  {journal} {\bibinfo  {journal}
			{Nature}\ }\textbf {\bibinfo {volume} {440}},\ \bibinfo {pages} {900}
		(\bibinfo {year} {2006})}\BibitemShut {NoStop}%
	\bibitem [{\citenamefont {Kr\"{u}ger}\ \emph {et~al.}(2010)\citenamefont
		{Kr\"{u}ger}, \citenamefont {Hofferberth}, \citenamefont {Mazets},
		\citenamefont {Lesanovsky},\ and\ \citenamefont
		{Schmiedmayer}}]{kruger_weakly_2010}%
	\BibitemOpen
	\bibfield  {author} {\bibinfo {author} {\bibfnamefont {P.}~\bibnamefont
			{Kr\"{u}ger}}, \bibinfo {author} {\bibfnamefont {S.}~\bibnamefont
			{Hofferberth}}, \bibinfo {author} {\bibfnamefont {I.~E.}\ \bibnamefont
			{Mazets}}, \bibinfo {author} {\bibfnamefont {I.}~\bibnamefont {Lesanovsky}},
		\ and\ \bibinfo {author} {\bibfnamefont {J.}~\bibnamefont {Schmiedmayer}},\
	}\href {\doibase 10.1103/PhysRevLett.105.265302} {\bibfield  {journal}
		{\bibinfo  {journal} {Phys. Rev. Lett.}\ }\textbf {\bibinfo {volume} {105}},\
		\bibinfo {pages} {265302} (\bibinfo {year} {2010})}\BibitemShut {NoStop}%
	\bibitem [{\citenamefont {Cazalilla}\ \emph {et~al.}(2011)\citenamefont
		{Cazalilla}, \citenamefont {Citro}, \citenamefont {Giamarchi}, \citenamefont
		{Orignac},\ and\ \citenamefont {Rigol}}]{cazalilla_one_2011}%
	\BibitemOpen
	\bibfield  {author} {\bibinfo {author} {\bibfnamefont {M.~A.}\ \bibnamefont
			{Cazalilla}}, \bibinfo {author} {\bibfnamefont {R.}~\bibnamefont {Citro}},
		\bibinfo {author} {\bibfnamefont {T.}~\bibnamefont {Giamarchi}}, \bibinfo
		{author} {\bibfnamefont {E.}~\bibnamefont {Orignac}}, \ and\ \bibinfo
		{author} {\bibfnamefont {M.}~\bibnamefont {Rigol}},\ }\href {\doibase
		10.1103/RevModPhys.83.1405} {\bibfield  {journal} {\bibinfo  {journal} {Rev.
				Mod. Phys.}\ }\textbf {\bibinfo {volume} {83}},\ \bibinfo {pages} {1405}
		(\bibinfo {year} {2011})}\BibitemShut {NoStop}%
	\bibitem [{\citenamefont {Guan}\ \emph {et~al.}(2013)\citenamefont {Guan},
		\citenamefont {Batchelor},\ and\ \citenamefont {Lee}}]{guan_fermi_2013}%
	\BibitemOpen
	\bibfield  {author} {\bibinfo {author} {\bibfnamefont {X.-W.}\ \bibnamefont
			{Guan}}, \bibinfo {author} {\bibfnamefont {M.~T.}\ \bibnamefont {Batchelor}},
		\ and\ \bibinfo {author} {\bibfnamefont {C.}~\bibnamefont {Lee}},\ }\href
	{\doibase 10.1103/RevModPhys.85.1633} {\bibfield  {journal} {\bibinfo
			{journal} {Rev. Mod. Phys.}\ }\textbf {\bibinfo {volume} {85}},\ \bibinfo
		{pages} {1633} (\bibinfo {year} {2013})}\BibitemShut {NoStop}%
	\bibitem [{\citenamefont {Giamarchi}(2003)}]{Giamarchi}%
	\BibitemOpen
	\bibfield  {author} {\bibinfo {author} {\bibfnamefont {T.}~\bibnamefont
			{Giamarchi}},\ }\href@noop {} {\emph {\bibinfo {title} {Quantum Physics in
				One Dimension}}}\ (\bibinfo  {publisher} {Clarendon Press, Oxford},\ \bibinfo
	{year} {2003})\BibitemShut {NoStop}%
	\bibitem [{\citenamefont {Haldane}(1981)}]{haldane_effective_1981}%
	\BibitemOpen
	\bibfield  {author} {\bibinfo {author} {\bibfnamefont {F.~D.~M.}\
			\bibnamefont {Haldane}},\ }\href {\doibase 10.1103/PhysRevLett.47.1840}
	{\bibfield  {journal} {\bibinfo  {journal} {Phys. Rev. Lett.}\ }\textbf
		{\bibinfo {volume} {47}},\ \bibinfo {pages} {1840} (\bibinfo {year}
		{1981})}\BibitemShut {NoStop}%
	\bibitem [{\citenamefont {Imambekov}\ \emph {et~al.}(2012)\citenamefont
		{Imambekov}, \citenamefont {Schmidt},\ and\ \citenamefont
		{Glazman}}]{imambekov_one-dimensional_2012}%
	\BibitemOpen
	\bibfield  {author} {\bibinfo {author} {\bibfnamefont {A.}~\bibnamefont
			{Imambekov}}, \bibinfo {author} {\bibfnamefont {T.~L.}\ \bibnamefont
			{Schmidt}}, \ and\ \bibinfo {author} {\bibfnamefont {L.~I.}\ \bibnamefont
			{Glazman}},\ }\href {\doibase 10.1103/RevModPhys.84.1253} {\bibfield
		{journal} {\bibinfo  {journal} {Rev. Mod. Phys.}\ }\textbf {\bibinfo {volume}
			{84}},\ \bibinfo {pages} {1253} (\bibinfo {year} {2012})}\BibitemShut
	{NoStop}%
	\bibitem [{\citenamefont {Rozhkov}(2005)}]{rozhkov_fermionic_2005}%
	\BibitemOpen
	\bibfield  {author} {\bibinfo {author} {\bibfnamefont {A.~V.}\ \bibnamefont
			{Rozhkov}},\ }\href {\doibase 10.1140/epjb/e2005-00312-3} {\bibfield
		{journal} {\bibinfo  {journal} {Eur. Phys. J. B}\ }\textbf {\bibinfo {volume}
			{47}},\ \bibinfo {pages} {193} (\bibinfo {year} {2005})}\BibitemShut
	{NoStop}%
	\bibitem [{\citenamefont {Lieb}(1963)}]{lieb_exact_1963b}%
	\BibitemOpen
	\bibfield  {author} {\bibinfo {author} {\bibfnamefont {E.~H.}\ \bibnamefont
			{Lieb}},\ }\href {\doibase 10.1103/PhysRev.130.1616} {\bibfield  {journal}
		{\bibinfo  {journal} {Phys. Rev.}\ }\textbf {\bibinfo {volume} {130}},\
		\bibinfo {pages} {1616} (\bibinfo {year} {1963})}\BibitemShut {NoStop}%
	\bibitem [{\citenamefont {Pines}\ and\ \citenamefont
		{Nozi\`{e}res}(1989)}]{pines+nozieres}%
	\BibitemOpen
	\bibfield  {author} {\bibinfo {author} {\bibfnamefont {D.}~\bibnamefont
			{Pines}}\ and\ \bibinfo {author} {\bibfnamefont {P.}~\bibnamefont
			{Nozi\`{e}res}},\ }\href@noop {} {\emph {\bibinfo {title} {The Theory of
				Quantum Liquids}}}\ (\bibinfo  {publisher} {Westview Press, Boulder},\
	\bibinfo {year} {1989})\BibitemShut {NoStop}%
	\bibitem [{\citenamefont {Pereira}\ \emph {et~al.}(2006)\citenamefont
		{Pereira}, \citenamefont {Sirker}, \citenamefont {Caux}, \citenamefont
		{Hagemans}, \citenamefont {Maillet}, \citenamefont {White},\ and\
		\citenamefont {Affleck}}]{pereira_dynamical_2006}%
	\BibitemOpen
	\bibfield  {author} {\bibinfo {author} {\bibfnamefont {R.~G.}\ \bibnamefont
			{Pereira}}, \bibinfo {author} {\bibfnamefont {J.}~\bibnamefont {Sirker}},
		\bibinfo {author} {\bibfnamefont {J.-S.}\ \bibnamefont {Caux}}, \bibinfo
		{author} {\bibfnamefont {R.}~\bibnamefont {Hagemans}}, \bibinfo {author}
		{\bibfnamefont {J.~M.}\ \bibnamefont {Maillet}}, \bibinfo {author}
		{\bibfnamefont {S.~R.}\ \bibnamefont {White}}, \ and\ \bibinfo {author}
		{\bibfnamefont {I.}~\bibnamefont {Affleck}},\ }\href {\doibase
		10.1103/PhysRevLett.96.257202} {\bibfield  {journal} {\bibinfo  {journal}
			{Phys. Rev. Lett.}\ }\textbf {\bibinfo {volume} {96}},\ \bibinfo {pages}
		{257202} (\bibinfo {year} {2006})}\BibitemShut {NoStop}%
	\bibitem [{\citenamefont {Matveev}\ and\ \citenamefont
		{Pustilnik}(2016)}]{matveev_effective_2016}%
	\BibitemOpen
	\bibfield  {author} {\bibinfo {author} {\bibfnamefont {K.~A.}\ \bibnamefont
			{Matveev}}\ and\ \bibinfo {author} {\bibfnamefont {M.}~\bibnamefont
			{Pustilnik}},\ }\href {\doibase 10.1103/PhysRevB.94.115436} {\bibfield
		{journal} {\bibinfo  {journal} {Phys. Rev. B}\ }\textbf {\bibinfo {volume}
			{94}},\ \bibinfo {pages} {115436} (\bibinfo {year} {2016})}\BibitemShut
	{NoStop}%
	\bibitem [{\citenamefont {Khodas}\ \emph {et~al.}(2007)\citenamefont {Khodas},
		\citenamefont {Pustilnik}, \citenamefont {Kamenev},\ and\ \citenamefont
		{Glazman}}]{khodas_dynamics_2007}%
	\BibitemOpen
	\bibfield  {author} {\bibinfo {author} {\bibfnamefont {M.}~\bibnamefont
			{Khodas}}, \bibinfo {author} {\bibfnamefont {M.}~\bibnamefont {Pustilnik}},
		\bibinfo {author} {\bibfnamefont {A.}~\bibnamefont {Kamenev}}, \ and\
		\bibinfo {author} {\bibfnamefont {L.~I.}\ \bibnamefont {Glazman}},\ }\href
	{\doibase 10.1103/PhysRevLett.99.110405} {\bibfield  {journal} {\bibinfo
			{journal} {Phys. Rev. Lett.}\ }\textbf {\bibinfo {volume} {99}},\ \bibinfo
		{pages} {110405} (\bibinfo {year} {2007})}\BibitemShut {NoStop}%
	\bibitem [{\citenamefont {Pereira}\ \emph {et~al.}(2008)\citenamefont
		{Pereira}, \citenamefont {White},\ and\ \citenamefont
		{Affleck}}]{pereira_exact_2008}%
	\BibitemOpen
	\bibfield  {author} {\bibinfo {author} {\bibfnamefont {R.~G.}\ \bibnamefont
			{Pereira}}, \bibinfo {author} {\bibfnamefont {S.~R.}\ \bibnamefont {White}},
		\ and\ \bibinfo {author} {\bibfnamefont {I.}~\bibnamefont {Affleck}},\ }\href
	{\doibase 10.1103/PhysRevLett.100.027206} {\bibfield  {journal} {\bibinfo
			{journal} {Phys. Rev. Lett.}\ }\textbf {\bibinfo {volume} {100}},\ \bibinfo
		{pages} {027206} (\bibinfo {year} {2008})}\BibitemShut {NoStop}%
	\bibitem [{\citenamefont {Imambekov}\ and\ \citenamefont
		{Glazman}(2008)}]{imambekov_exact_2008}%
	\BibitemOpen
	\bibfield  {author} {\bibinfo {author} {\bibfnamefont {A.}~\bibnamefont
			{Imambekov}}\ and\ \bibinfo {author} {\bibfnamefont {L.~I.}\ \bibnamefont
			{Glazman}},\ }\href {\doibase 10.1103/PhysRevLett.100.206805} {\bibfield
		{journal} {\bibinfo  {journal} {Phys. Rev. Lett.}\ }\textbf {\bibinfo
			{volume} {100}},\ \bibinfo {pages} {206805} (\bibinfo {year}
		{2008})}\BibitemShut {NoStop}%
	\bibitem [{\citenamefont {Cheianov}\ and\ \citenamefont
		{Pustilnik}(2008)}]{cheianov_threshold_2008}%
	\BibitemOpen
	\bibfield  {author} {\bibinfo {author} {\bibfnamefont {V.~V.}\ \bibnamefont
			{Cheianov}}\ and\ \bibinfo {author} {\bibfnamefont {M.}~\bibnamefont
			{Pustilnik}},\ }\href {\doibase 10.1103/PhysRevLett.100.126403} {\bibfield
		{journal} {\bibinfo  {journal} {Phys. Rev. Lett.}\ }\textbf {\bibinfo
			{volume} {100}},\ \bibinfo {pages} {126403} (\bibinfo {year}
		{2008})}\BibitemShut {NoStop}%
	\bibitem [{\citenamefont {Matveev}\ and\ \citenamefont
		{Furusaki}(2008)}]{matveev_spectral_2008}%
	\BibitemOpen
	\bibfield  {author} {\bibinfo {author} {\bibfnamefont {K.~A.}\ \bibnamefont
			{Matveev}}\ and\ \bibinfo {author} {\bibfnamefont {A.}~\bibnamefont
			{Furusaki}},\ }\href {\doibase 10.1103/PhysRevLett.101.170403} {\bibfield
		{journal} {\bibinfo  {journal} {Phys. Rev. Lett.}\ }\textbf {\bibinfo
			{volume} {101}},\ \bibinfo {pages} {170403} (\bibinfo {year}
		{2008})}\BibitemShut {NoStop}%
	\bibitem [{\citenamefont {Imambekov}\ and\ \citenamefont
		{Glazman}(2009)}]{imambekov_phenomenology_2009}%
	\BibitemOpen
	\bibfield  {author} {\bibinfo {author} {\bibfnamefont {A.}~\bibnamefont
			{Imambekov}}\ and\ \bibinfo {author} {\bibfnamefont {L.~I.}\ \bibnamefont
			{Glazman}},\ }\href {\doibase 10.1103/PhysRevLett.102.126405} {\bibfield
		{journal} {\bibinfo  {journal} {Phys. Rev. Lett.}\ }\textbf {\bibinfo
			{volume} {102}},\ \bibinfo {pages} {126405} (\bibinfo {year}
		{2009})}\BibitemShut {NoStop}%
	\bibitem [{\citenamefont {Pereira}\ \emph {et~al.}(2009)\citenamefont
		{Pereira}, \citenamefont {White},\ and\ \citenamefont
		{Affleck}}]{pereira_spectral_2009}%
	\BibitemOpen
	\bibfield  {author} {\bibinfo {author} {\bibfnamefont {R.~G.}\ \bibnamefont
			{Pereira}}, \bibinfo {author} {\bibfnamefont {S.~R.}\ \bibnamefont {White}},
		\ and\ \bibinfo {author} {\bibfnamefont {I.}~\bibnamefont {Affleck}},\ }\href
	{\doibase 10.1103/PhysRevB.79.165113} {\bibfield  {journal} {\bibinfo
			{journal} {Phys. Rev. B}\ }\textbf {\bibinfo {volume} {79}},\ \bibinfo
		{pages} {165113} (\bibinfo {year} {2009})}\BibitemShut {NoStop}%
	\bibitem [{\citenamefont {Zvonarev}\ \emph {et~al.}(2009)\citenamefont
		{Zvonarev}, \citenamefont {Cheianov},\ and\ \citenamefont
		{Giamarchi}}]{zvonarev_edge_2009}%
	\BibitemOpen
	\bibfield  {author} {\bibinfo {author} {\bibfnamefont {M.~B.}\ \bibnamefont
			{Zvonarev}}, \bibinfo {author} {\bibfnamefont {V.~V.}\ \bibnamefont
			{Cheianov}}, \ and\ \bibinfo {author} {\bibfnamefont {T.}~\bibnamefont
			{Giamarchi}},\ }\href {\doibase 10.1103/PhysRevB.80.201102} {\bibfield
		{journal} {\bibinfo  {journal} {Phys. Rev. B}\ }\textbf {\bibinfo {volume}
			{80}},\ \bibinfo {pages} {201102} (\bibinfo {year} {2009})}\BibitemShut
	{NoStop}%
	\bibitem [{\citenamefont {Kamenev}\ and\ \citenamefont
		{Glazman}(2009)}]{kamenev_dynamics_2009}%
	\BibitemOpen
	\bibfield  {author} {\bibinfo {author} {\bibfnamefont {A.}~\bibnamefont
			{Kamenev}}\ and\ \bibinfo {author} {\bibfnamefont {L.~I.}\ \bibnamefont
			{Glazman}},\ }\href {\doibase 10.1103/PhysRevA.80.011603} {\bibfield
		{journal} {\bibinfo  {journal} {Phys. Rev. A}\ }\textbf {\bibinfo {volume}
			{80}},\ \bibinfo {pages} {011603} (\bibinfo {year} {2009})}\BibitemShut
	{NoStop}%
	\bibitem [{\citenamefont {Shashi}\ \emph {et~al.}(2012)\citenamefont {Shashi},
		\citenamefont {Panfil}, \citenamefont {Caux},\ and\ \citenamefont
		{Imambekov}}]{shashi_exact_2012}%
	\BibitemOpen
	\bibfield  {author} {\bibinfo {author} {\bibfnamefont {A.}~\bibnamefont
			{Shashi}}, \bibinfo {author} {\bibfnamefont {M.}~\bibnamefont {Panfil}},
		\bibinfo {author} {\bibfnamefont {J.-S.}\ \bibnamefont {Caux}}, \ and\
		\bibinfo {author} {\bibfnamefont {A.}~\bibnamefont {Imambekov}},\ }\href
	{\doibase 10.1103/PhysRevB.85.155136} {\bibfield  {journal} {\bibinfo
			{journal} {Phys. Rev. B}\ }\textbf {\bibinfo {volume} {85}},\ \bibinfo
		{pages} {155136} (\bibinfo {year} {2012})}\BibitemShut {NoStop}%
	\bibitem [{\citenamefont {Meinert}\ \emph {et~al.}(2015)\citenamefont
		{Meinert}, \citenamefont {Panfil}, \citenamefont {Mark}, \citenamefont
		{Lauber}, \citenamefont {Caux},\ and\ \citenamefont
		{N\"{a}gerl}}]{meinert_probing_2015}%
	\BibitemOpen
	\bibfield  {author} {\bibinfo {author} {\bibfnamefont {F.}~\bibnamefont
			{Meinert}}, \bibinfo {author} {\bibfnamefont {M.}~\bibnamefont {Panfil}},
		\bibinfo {author} {\bibfnamefont {M.~J.}~\bibnamefont {Mark}}, \bibinfo {author}
		{\bibfnamefont {K.}~\bibnamefont {Lauber}}, \bibinfo {author} {\bibfnamefont
			{J.-S.}\ \bibnamefont {Caux}}, \ and\ \bibinfo {author} {\bibfnamefont
			{H.-C.}\ \bibnamefont {N\"{a}gerl}},\ }\href {\doibase
		10.1103/PhysRevLett.115.085301} {\bibfield  {journal} {\bibinfo  {journal}
			{Phys. Rev. Lett.}\ }\textbf {\bibinfo {volume} {115}},\ \bibinfo {pages}
		{085301} (\bibinfo {year} {2015})}\BibitemShut {NoStop}%
	\bibitem [{\citenamefont {Fabbri}\ \emph {et~al.}(2015)\citenamefont {Fabbri},
		\citenamefont {Panfil}, \citenamefont {Cl\'{e}ment}, \citenamefont {Fallani},
		\citenamefont {Inguscio}, \citenamefont {Fort},\ and\ \citenamefont
		{Caux}}]{fabbri_dynamical_2015}%
	\BibitemOpen
	\bibfield  {author} {\bibinfo {author} {\bibfnamefont {N.}~\bibnamefont
			{Fabbri}}, \bibinfo {author} {\bibfnamefont {M.}~\bibnamefont {Panfil}},
		\bibinfo {author} {\bibfnamefont {D.}~\bibnamefont {Cl\'{e}ment}}, \bibinfo
		{author} {\bibfnamefont {L.}~\bibnamefont {Fallani}}, \bibinfo {author}
		{\bibfnamefont {M.}~\bibnamefont {Inguscio}}, \bibinfo {author}
		{\bibfnamefont {C.}~\bibnamefont {Fort}}, \ and\ \bibinfo {author}
		{\bibfnamefont {J.-S.}\ \bibnamefont {Caux}},\ }\href {\doibase
		10.1103/PhysRevA.91.043617} {\bibfield  {journal} {\bibinfo  {journal} {Phys.
				Rev. A}\ }\textbf {\bibinfo {volume} {91}},\ \bibinfo {pages} {043617}
		(\bibinfo {year} {2015})}\BibitemShut {NoStop}%
	\bibitem [{\citenamefont {Lieb}\ and\ \citenamefont
		{Liniger}(1963)}]{lieb_exact_1963}%
	\BibitemOpen
	\bibfield  {author} {\bibinfo {author} {\bibfnamefont {E.~H.}\ \bibnamefont
			{Lieb}}\ and\ \bibinfo {author} {\bibfnamefont {W.}~\bibnamefont {Liniger}},\
	}\href {\doibase 10.1103/PhysRev.130.1605} {\bibfield  {journal} {\bibinfo
			{journal} {Phys. Rev.}\ }\textbf {\bibinfo {volume} {130}},\ \bibinfo {pages}
		{1605} (\bibinfo {year} {1963})}\BibitemShut {NoStop}%
	\bibitem [{\citenamefont {Sutherland}(2004)}]{sutherland}%
	\BibitemOpen
	\bibfield  {author} {\bibinfo {author} {\bibfnamefont {B.}~\bibnamefont
			{Sutherland}},\ }\href@noop {} {\emph {\bibinfo {title} {Beautiful models}}}\
	(\bibinfo  {publisher} {World Scientific, Singapore},\ \bibinfo {year}
	{2004})\BibitemShut {NoStop}%
	\bibitem [{\citenamefont {Korepin}\ \emph {et~al.}(1993)\citenamefont
		{Korepin}, \citenamefont {Bogoliubov},\ and\ \citenamefont
		{Izergin}}]{korepin1993book}%
	\BibitemOpen
	\bibfield  {author} {\bibinfo {author} {\bibfnamefont {V.~E.}\ \bibnamefont
			{Korepin}}, \bibinfo {author} {\bibfnamefont {N.~M.}\ \bibnamefont
			{Bogoliubov}}, \ and\ \bibinfo {author} {\bibfnamefont {A.~G.}\ \bibnamefont
			{Izergin}},\ }\href@noop {} {\emph {\bibinfo {title} {Quantum inverse
				scattering method and correlation functions}}}\ (\bibinfo  {publisher}
	{Cambridge University Press},\ \bibinfo {year} {1993})\BibitemShut {NoStop}%
	\bibitem [{Note1()}]{Note1}%
	\BibitemOpen
	\bibinfo {note} {The notable exception where the density of rapidities is
		discontinuous is the integrable model with the inverse-square potential,
		which has a singular two-particle phase shift}\BibitemShut {NoStop}%
	\bibitem [{\citenamefont {Popov}(1977)}]{popov_theory_1977}%
	\BibitemOpen
	\bibfield  {author} {\bibinfo {author} {\bibfnamefont {V.~N.}\ \bibnamefont
			{Popov}},\ }\href {\doibase 10.1007/BF01036714} {\bibfield  {journal}
		{\bibinfo  {journal} {Theor. Math. Phys.}\ }\textbf {\bibinfo {volume}
			{30}},\ \bibinfo {pages} {222} (\bibinfo {year} {1977})}\BibitemShut
	{NoStop}%
	\bibitem [{\citenamefont {Tracy}\ and\ \citenamefont
		{Widom}(2016)}]{tracy_ground_2016}%
	\BibitemOpen
	\bibfield  {author} {\bibinfo {author} {\bibfnamefont {C.~A.}\ \bibnamefont
			{Tracy}}\ and\ \bibinfo {author} {\bibfnamefont {H.}~\bibnamefont {Widom}},\
	}\href {\doibase 10.1088/1751-8113/49/29/294001} {\bibfield  {journal}
		{\bibinfo  {journal} {J. Phys. A}\ }\textbf {\bibinfo {volume}
			{49}},\ \bibinfo {pages} {294001} (\bibinfo {year} {2016})}\BibitemShut
	{NoStop}%
	\bibitem [{\citenamefont {Prolhac}(2017)}]{prolhac_ground_2017}%
	\BibitemOpen
	\bibfield  {author} {\bibinfo {author} {\bibfnamefont {S.}~\bibnamefont
			{Prolhac}},\ }\href {\doibase 10.1088/1751-8121/aa5e00} {\bibfield  {journal}
		{\bibinfo  {journal} {J. Phys. A}\ }\textbf {\bibinfo {volume}
			{50}},\ \bibinfo {pages} {144001} (\bibinfo {year} {2017})}\BibitemShut
	{NoStop}%
	\bibitem [{\citenamefont {Hutson}(1963)}]{hutson_circular_1963}%
	\BibitemOpen
	\bibfield  {author} {\bibinfo {author} {\bibfnamefont {V.}~\bibnamefont
			{Hutson}},\ }\href {\doibase 10.1017/S0305004100002152} {\bibfield  {journal}
		{\bibinfo  {journal} {Math. Proc. Cambridge Philos. Soc.}\ }\textbf {\bibinfo
			{volume} {59}},\ \bibinfo {pages} {211} (\bibinfo {year} {1963})}\BibitemShut
	{NoStop}%
	\bibitem [{\citenamefont {Pustilnik}\ and\ \citenamefont
		{Matveev}(2014)}]{pustilnik_low-energy_2014}%
	\BibitemOpen
	\bibfield  {author} {\bibinfo {author} {\bibfnamefont {M.}~\bibnamefont
			{Pustilnik}}\ and\ \bibinfo {author} {\bibfnamefont {K.~A.}\ \bibnamefont
			{Matveev}},\ }\href {\doibase 10.1103/PhysRevB.89.100504} {\bibfield
		{journal} {\bibinfo  {journal} {Phys. Rev. B}\ }\textbf {\bibinfo {volume}
			{89}},\ \bibinfo {pages} {100504} (\bibinfo {year} {2014})}\BibitemShut
	{NoStop}%
	\bibitem [{\citenamefont {Kulish}\ \emph {et~al.}(1976)\citenamefont {Kulish},
		\citenamefont {Manakov},\ and\ \citenamefont
		{Faddeev}}]{kulish_comparison_1976}%
	\BibitemOpen
	\bibfield  {author} {\bibinfo {author} {\bibfnamefont {P.~P.}\ \bibnamefont
			{Kulish}}, \bibinfo {author} {\bibfnamefont {S.~V.}\ \bibnamefont {Manakov}},
		\ and\ \bibinfo {author} {\bibfnamefont {L.~D.}\ \bibnamefont {Faddeev}},\
	}\href {\doibase 10.1007/BF01028912} {\bibfield  {journal} {\bibinfo
			{journal} {Theor. Math. Phys.}\ }\textbf {\bibinfo {volume} {28}},\ \bibinfo
		{pages} {615} (\bibinfo {year} {1976})}\BibitemShut {NoStop}%
	\bibitem [{\citenamefont {Ristivojevic}(2014)}]{ristivojevic_excitation_2014}%
	\BibitemOpen
	\bibfield  {author} {\bibinfo {author} {\bibfnamefont {Z.}~\bibnamefont
			{Ristivojevic}},\ }\href {\doibase 10.1103/PhysRevLett.113.015301} {\bibfield
		{journal} {\bibinfo  {journal} {Phys. Rev. Lett.}\ }\textbf {\bibinfo
			{volume} {113}},\ \bibinfo {pages} {015301} (\bibinfo {year}
		{2014})}\BibitemShut {NoStop}%
	\bibitem [{\citenamefont {Lang}\ \emph {et~al.}(2017)\citenamefont {Lang},
		\citenamefont {Hekking},\ and\ \citenamefont
		{Minguzzi}}]{lang_ground-state_2017}%
	\BibitemOpen
	\bibfield  {author} {\bibinfo {author} {\bibfnamefont {G.}~\bibnamefont
			{Lang}}, \bibinfo {author} {\bibfnamefont {F.}~\bibnamefont {Hekking}}, \
		and\ \bibinfo {author} {\bibfnamefont {A.}~\bibnamefont {Minguzzi}},\ }\href
	{\doibase 10.21468/SciPostPhys.3.1.003} {\bibfield  {journal} {\bibinfo
			{journal} {SciPost Phys.}\ }\textbf {\bibinfo {volume} {3}},\ \bibinfo
		{pages} {003} (\bibinfo {year} {2017})}\BibitemShut {NoStop}%
\end{thebibliography}
%

\end{document}